\documentclass{emulateapj}
\def\xtej{XTE~J1739-285}
\usepackage{graphicx,subfigure}
\begin{document}
\title{Constraints to the EOS of ultradense
  matter with model-independent astrophysical observations.}
\author{G. Lavagetto\altaffilmark{1},  
I. Bombaci\altaffilmark{2}, A. D'A\`\i\altaffilmark{1}, I. Vida\~na\altaffilmark{3} and N. R. Robba\altaffilmark{1}}
\altaffiltext{1}{ Dipartimento di Scienze Fisiche ed Astronomiche, 
Universit\`a di Palermo, via Archirafi n.36, 90123 Palermo, Italy.}
\altaffiltext{2}{Dipartimento di Fisica, Universit\`a di Pisa, and INFN sezione di Pisa,  
largo B. Pontecorvo, n. 3, 56127, Pisa, Italy}
\altaffiltext{3}{Departament d'Estructura i Costituents de la Mat\`eria, 
Universitat de Barcelona, E-08028 Barcelona, Spain}
\label{firstpage}
\begin{abstract}
The recent discovery of burst oscillations at 1122 Hz in the x-ray
transient \xtej, together with the measurement of the mass of the
binary millisecond pulsar PSR~J0751+1807 ($2.1\pm 0.2~M_\odot$) 
can finally allow us to put strong, model-independent observational
constraints to the equation of state of compact stars. 
We show that the measurement of the moment of inertia of
PSR~J0737+3039A, together with these constraints, could allow to
discriminate further the details of the inner structure of neutron stars.
Moreover, we show that if \xtej\, is constituted of nucleonic matter,
 any equation of state allows only a narrow range of very high masses, and this
 could explain why up to now compact stars spinning faster than a 
 millisecond have been so difficult to detect.
\end{abstract}
\keywords{
Stars: neutron -- X-rays: binaries -- binaries: close -- relativity.}

\section{Introduction}

Ever since the discovery of the first neutron star (NS) by \citet{heal68},  
these compact objects have been regarded as the ideal 
experimental test-bed for the theories of the state of matter at 
supernuclear densities.  In particular, the mass, the radius, the 
attainable spin frequency of a NS strongly depend on the 
equation of state (EOS) governing matter at these densities. 

Different models for the EOS of dense hadronic matter  predict  a NS  
maximum mass ($M_\mathrm{max}$) in the range of   1.4--2.6 $M_\odot$,  and 
a  corresponding central density $n_c$  in range of 4--10 times the saturation 
density  ($n_0 = 0.16~ {\rm fm}^{-3}$)  of nuclear matter \citep{prak97}.    
In the case  of a star with $ M \sim 1.4~M_\odot$, different EOS models predict 
a radius  in the range of 7--16 km.     

Due to the large value of the stellar central densities, various regimes  
(particle species and phases) of dense hadronic matter are expected 
in the interiors of NS.  Consequently, different types of ``neutron stars''  
(compact stars, CSs)  are hypothesized.  
In the simplest model  the core of a neutron star  is described  as a uniform fluid 
of neutron rich nuclear matter in equilibrium with respect 
to the weak interaction: these are the so-called ``traditional'' NSs.    
However, due to the large value of the stellar central density and to the 
rapid increase of the nucleon chemical potentials with density,   
hyperons ($\Lambda$, $\Sigma^{-}$, $\Sigma^{0}$, $\Sigma^{+}$, $\Xi^{-}$ 
and $\Xi^{0}$ particles) are expected to appear in the inner core of the star.    
Other {\it exotic} phases of hadronic matter such as  a Bose-Einstein condensate 
of negative pion ($\pi^-$) or negative kaon ($K^-$)  could be present in 
the inner part of the star.  
The core of the more massive NS is also one of the best candidates 
in the Universe for a  phase transition from hadronic matter to a 
deconfined quark phase to occur.  
Compact stars  which possess a ``quark matter core'' either as a 
mixed phase of deconfined quarks and hadrons or as a pure quark matter (QM) 
phase are called {\it Hybrid Stars} \citep[HyS,][]{gle96}.     
The  more {\it conventional} NSs  in which no fraction of QM is present, 
are referred to as  {\it pure Hadronic Stars} (HSs).  

Even more challenging than the existence of a quark core in a NS, 
is the possible existence of a new type  of CS consisting completely 
of a charge neutral deconfined  mixture of {\it up},  {\it down}, {\it strange } quarks 
and electrons,  satisfying the hypothesis on the absolute stability \citep{bodm,witt} 
of strange quark matter (SQM).   
Such CSs have been called {\it strange stars} (SS).     
The analysis of different type of observational  data has given indirect evidence 
for the possible existence of  SS (see {\it e.g.} \citep{li99a}).  
We will refer to hybrid stars and strange stars  collectively as {\it Quark Stars} (QS).   

Unluckily, for decades no significant constraint on both the mass and
the radius of a single object could be extracted from
astrophysical observations. While the determination of the masses of some CSs 
was made possible thanks to the study of the timing motion of radio pulsars in binary
systems, the radius determination, or alternatively the moment of inertia, has
been more elusive. Today the measured masses of CSs lie around the average
value of 1.35 $\pm$ 0.04 $M_{\odot}$  and are roughly distributed according 
to a Gaussian distribution \citep{tocha99}. 

In recent years the discovery of kilohertz quasi-periodic oscillations (QPOs)
in the power spectra of several accreting CSs has opened new possibilities,
as they are phenomena whose timescale corresponds to the dynamical timescale 
in the very neighborhood of a CSs \citep[see][for a review]{vdk06}. 
These QPOs have peak frequencies in the 200-1200 Hz range and usually  
appear as two broad peaks (denoted upper and lower kHz QPO), that simultaneously move up and down
according to the source accretion state. The lower kHz QPO sometimes becomes undetected as the
upper kHz QPO reaches lower frequencies; for low accretion rates both peaks can disappear.
While their peak frequencies smoothly change with the accretion rate
of the source, their peak separation is roughly constant and equal, within 20\%, to the
spin frequency of the NS, or, for some sources, to the half of it \citep{wijal03}. 
The reason for this discrepancy is still unaccounted for.
Most theoretical models of kHz QPOs associate the frequency of the highest peak
to the motion of matter at the last stable orbit of an accretion disk around the CS. 
If this is true, assuming that matter in the accretion disc is in
Keplerian motion, it is possible to derive an upper limit to the density of the
CS. This, in turn,  puts some limits to the EOS of CSs \citep[see for
example][]{zaal06}, but still relies on a theoretical interpretation of 
a phenomenon which is not completely understood.

Another observational constraint comes from the reported gravitational redshift of 
absorption lines produced in the photosphere of the X-ray binary system EXO 0748-676
\citep*{coal02}. The redshift ($z$ = 0.35) implies a
compactness parameter for the CS of $ GM/Rc^2 \simeq 0.23$. 
However, even this result (which anyway is not able 
to constrain the EOS significantly if taken alone) is dependent both
on the model adopted for the spectral fitting and on the assumptions on the
emitting region. 

Another astrophysical observable is the radiation radius $R_\infty$ of
the emission from isolated CSs: the nearest and better studied
isolated CS, RX~J1856.5-3754, has a spectrum that can be well
fitted with a blackbody of 57 eV temperature. This value 
sets an upper limit for the $R_\infty$ of this object,
which was estimated $\leq$ 8 km \citep{drake}.
This limit, which is quite demanding, can imply an EOS based
on QM. However, the determination of this radius is strongly dependent
on the distance of the source and on theoretical issues regarding
the geometry of the emission  and the composition of the atmosphere.
\citet{trumper04} has recently argued against the constraints found by
\citet{drake}, claiming that the upper limit is much higher. 

The situation is even worse if we try to
infer the radius of the CS from the blackbody temperature of an accreting
X-ray binary \citep[see e.g.][]{zaal06}: in this case even the origin of the
emission can be debatable, as the blackbody emission can be originated
near the stellar surface as well as from the accretion disc. 

A truly model-independent constraint has been found with the
measurement of the mass of several millisecond radio pulsars thanks to the
detection of post-Newtonian corrections to their orbital motion. In
particular, \citet{nial05} found that PSR J0751+1807 has a mass of 
$2.1 \pm 0.2 M_\odot$ (1 $\sigma$ error), and \citet{raal05} found that pulsar 
I of the globular cluster Terzan 5 has a mass that exceeds 1.68 $M_\odot$ at
95\% confidence level, showing that CSs can have quite large masses.

Another model independent limit to the EOS at supernuclear densities comes from
laboratory experiments on heavy ions collisions (HIC): \citet{klaal06}
show that some constraints on the EOS can be given by flow data and
sub-threshold Kaon production in HIC. They found that these constraints
would rule out some of the very stiff EOSs. \citet*{grial06} solved the problem 
by simply assuming that above a certain
energy density, a quark core forms, with a phase transition to stiff,
color superconducting QM that can alter only the behavior of the EOS
at very high densities, making it softer near the maximum mass. In
practice, this results in small modifications to the behavior of the
EOS that do not alter very significantly the mass-radius relationship
of CSs.

Finally, we will show that the recent discovery of burst oscillations with a frequency
of 1122 Hz in \xtej\ \citep{karal06}, is potentially
another important constraint to the EOS of CSs: 
these oscillations, occurring during type I X-ray bursts, have the
same frequency of the spin frequency of the CS \citep{cha03}. This
discovery can put a real, model independent limit to the
radius of the NS in \xtej, and thus to its EOS.

This discovery rules out some speculations that were made in order to explain
the inability of CSs to spin up to periods
below one millisecond. It was argued that the lack of these 
fast spinning objects was due to the emission
of gravitational waves during accretion \citep*{aks98,bild98} or to
the disc-magnetic field interaction in low mass
X-ray binaries \citep[see e.g.][]{andal05}. All these theories have
been wiped out by this discovery, but the problem has now turned to: why only one
CS is observed to spin below one millisecond?

In this letter we first impose the model independent constraints to
the EOS of CSs, we discuss how future further observations can improve these constraints, 
 then we try to explain the difficulty in finding very
fast objects on the basis of our findings on the structure of CSs.

\section{Model-independent constraints}

\begin{figure}
  \centering 
  \includegraphics[scale=0.32, angle=0]{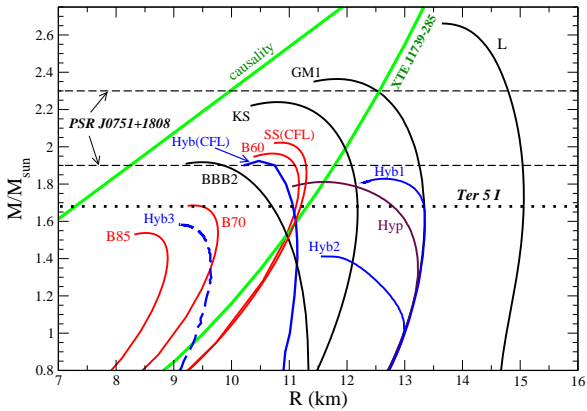}
  \caption{Mass-radius plane with mass limits (dashed black lines) and spin and causality limits (green lines), plus sequences for all EOSs considered (see text for details).}
  \label{fig:EOS}
\end{figure}

In figure \ref{fig:EOS} we show in the mass-radius plane the sequences 
for non-rotating  CSs with very different compositions,  together with 
constraints coming from model-independent observational results.
First, the sequence should include masses as large as the one of PSR J0751+1807 
(horizontal dashed lines) and of pulsar I of Terzan 5 (dotted line). 
We assume that the maximum non-rotating mass is almost equal to 
the maximum mass at the spin frequencies of these
objects, which is a good approximation given the relatively low spin
frequency of both objects. 
Second, the mass-radius relation should be such to allow the star to
spin up to the maximum spin rate observed to date, i.e. to 1122 Hz, as
observed in \xtej. 

The maximum rate of rotation ($\Omega_\mathrm{max}$) sustainable by a CS 
strongly depends on the overall stiffness of the EOS of dense hadronic matter. 
Equilibrium sequences of rapidly rotating CSs have been constructed 
numerically in general relativity by several groups  \citep*{cst94, datal98, bomal00}. 
The numerical results for $\Omega_\mathrm{max}$  obtained for a broad set of realistic EOS 
can be reproduced with a very good accuracy using simple empirical formulas 
which relate $\Omega_\mathrm{max}$ to the mass ($M_\mathrm{max}$) and radius $R_0$  of the 
non-rotating maximum mass configuration \citep{saal94}.
The following simple empirical formula given by Lattimer (Conference 
"Isolated Neutron Stars, London 2006, see also Latimer \& Prakash 2004)
approximately  describes the minimum rotation period  
$P_\mathrm{min} = 2 \pi/\Omega_\mathrm{max}$ for a star of mass $M$ and non-rotating radius $R$ 
\begin{equation}
 \label{eq:latt06}
      P_\mathrm{min} =0.96 \frac{(R/10 ~\mathrm{km})^{3/2}}{(M/M_\odot)^{1/2 }}~\mathrm{ms}.    
\end{equation}
Using the  rotation period $P = 0.891$~ms,  corresponding to the measured X-ray
burst oscillation frequency for \xtej, we get the following upper limit for the radius 
of the compact star in \xtej  
\begin{equation} 
  \label{eq:rlat}
  R< 9.52 \left(M/M_\odot\right)^{1/3}~\mathrm{km}.
\end{equation}
This condition is plotted in figure as a green line, with another
green line indicating the causality limit for the mass-radius
relation.

As we said before, there exist various possible classes of CSs. 
We considered several EOSs in order to match them with observations, 
and constructed non-rotating sequences varying the central density of the star. 
In figure \ref{fig:EOS} we report the mass-radius sequences for several EOSs. 
The curves relative to NSs  are colored in black,  
those for hyperon stars in brown. In blue we show the sequences for HySs, 
and in red those for SSs. 
In particular, we report results for the following classes of systems:  
i) pure neutron matter CSs (L) calculated within a relativistic field theoretical 
approach in the mean field approximation with meson exchange \citep{pasm75}.  
This outdated and  schematic model, is considered as an example of an 
extemely stiff EOS;   
ii) nucleonic stars (label BBB2)  built with an EOS calculated using the microscopic 
Brueckner-Hartree-Fock (BHF) many-body approach  with the Paris nucleon-nucleon 
interaction  plus  nuclear three-body forces \citep{bbb97};
iii) nucleonic stars (KS) built with an EOS  derived \citep{kraal06}
from the Dirac-BHF approach using the Bonn B nuclear interaction.  
iv) nucleonic stars (GM1) calculated within a relativistic field theoretical approach 
in the mean field approximation \citep{glemo91};
v) hyperon  stars (Hyp) calculated with the previous GM1  EOS with  
the inclusion of hyperons; 
vi) Hybrid star sequences calculated using the same  GM1 EOS  for hyperonic 
matter to describe the hadronic phase, and the MIT bag model EOS \citep{farhi} to describe the deconfined 
quark phase.  For the masses of the three quark flavors we took  
$m_u = m_d = 0$ and $m_s = 150$ MeV. 
For the value of the bag constant  we took $B = 208.24$~MeV/fm$^3$ (Hyb1) and 
$B = 136.63$~MeV/fm$^3$ (Hyb2).   
The curve (Hyb3) refers to hybrid stars constructed using a different 
parametrization for  the EOS by \citet{glemo91} (GM3)  for the hadronic phase 
and  using $B = 80$~MeV/fm$^3$; 
vii) Hybrid stars \citep[Hyb(CFL), ][]{alfal05}  
whose core is a mixed phase of nuclear matter and  color-superconducting quark 
phase in the so called "color flavor locked" (CFL) phase;   
viii) strange star sequences  (B85,  B70) built with the MIT bag model with 
 with $B = 85$~MeV/fm$^3$ and $B = 70$~MeV/fm$^3$
respectively ($m_u = m_d = 0$ and $m_s = 150$ MeV), and 
sequence (B60) with $B = 60$~MeV/fm$^3$ and massless quarks; 
ix) CFL strange stars  (label SS(CFL))  \citep{lug03}     
within the bag model EOS  with  $B = 70$~MeV/fm$^3$,    
$m_u = m_d = 0$ and $m_s = 150$ MeV and  quark pairing gap $\Delta = 100$ MeV.  

Looking at the figure we see that, although almost all classes of EOSs
are still allowed (apart probably from Hyperon stars), only the
stiffest SSs are allowed, and that while soft HSs are excluded by the
mass constraint, some
of the stiffest ones (i.e. Hyb1 and Hyb2, L) are excluded by the rotational
limit. Other EOSs are only marginally compatible with the constraints
(BBB2, GM1, B60 and CFL) and would require exceptional conditions 
(e.g. a mass almost equal to the limiting mass for gravitational
collapse) to meet the observational constraints.

\section{Discussion and Conclusions}

The exceptional discovery of a CS spinning with a period
below one millisecond allowed us to put a new model-independent
constraint to the EOS of matter above nuclear density. Combining this
with the constraint given by the highest masses measured in
millisecond radio pulsars, we are able to narrow the classes of EOSs
that could describe compact objects. As it is clear from figure
\ref{fig:EOS} we are left with two main classes of EOSs that comply
with our constraints: very stiff quark matter EOSs and some quite
stiff nucleonic matter EOSs. These two classes have completely
different characteristics.

SSs that satisfy
both constraints have radii smaller than $\sim 10.5$ km for
a mass of 1.337 M$_\odot$ (the mass of  the
ultrarelativistic pulsar PSR J737-303A). Using the formula
by \citet{bejhe02}, the moment of inertia of a SS with a mass of
1.4 M$_\odot$ and a radius of 10.5 km is, being  $x=M/R~(\mathrm{M_\odot/km})$
\begin{equation}
  \label{eq:inertia}
  I=\frac{2}{5}(1+x) M R^2 = 1.32 \times
  10^{45} ~\mathrm{g~cm^2}.
\end{equation}

 On the other hand, HSs that comply to both
constraints have radii in the range $12-13.5$ km for a mass of 1.4
M$_\odot$, and consequently larger moment of inertia: using the formula
by \citet{bejhe02} for HSs, assuming $R\simeq 13~\mathrm{km}$ we obtain
\begin{equation}
  \label{eq:inertia2}
  I=\frac{2}{9}(1+5x) M R^2 = 1.52 \times
  10^{45} ~\mathrm{g ~ cm^2}.
\end{equation}
There is a $13\%$ discrepancy between the two
values, so that the measurement of the moment of inertia of  PSR
J0737-3039A with the uncertainty of $10\%$ -
quoted as reachable by \citet{laschu05} - could tell us if this pulsar
is a SS or a very stiff
HS, although it will  not be possible to discriminate between a NS
and a HyS, as \citet{klaal06b} have shown. We have to note that if the
HS is one of the less stiff ones $R\simeq 11~\mathrm{km}$, its moment of
inertia would be virtually indistinguishable from the one of a stiff
SS. However, soft NSs (like EOS BBB2 of figure 1) are strongly
disfavored by the mass constraint. 

If we look at figure \ref{fig:EOS}, we can also notice that the
mass-radii sequences of the allowed HSs cross the spin limit line for \xtej\,
only at very high masses: as an example, the KS EOS allows only stars
exceeding $2~M_\odot$ to attain such high spins.
 This is because the radii of stiff HSs shrink only when
approaching the maximum allowed masses. So, not only
the moment of inertia of the HSs is large, meaning that it is
necessary to accrete more mass to get a star to periods in the order
of one millisecond, but also that only a very massive star would be
able to reach a period of $0.891~\mathrm{ms}$, and only at mass
shedding. Moreover, the HS should lie in a very narrow range of masses
in order to be stable against gravitational collapse: if the star is
supramassive (which can be the case of \xtej), accretion of further
matter can easily lead to the collapse to a 
black hole \citep{pa1}. 

On the contrary, since SSs are
self-bound objects, their radii expand with growing mass, allowing
very fast spins even for lower masses (i.e. EOS B60 can attain this
period for $M\ge 1.5 M_\odot$). This means that it is much
easier to attain the period of \xtej\, in this case.

Paradoxically, if - as the
discovery of \xtej\, implies - there is no
mechanism avoiding the spin up to millisecond periods \citep[see
e.g.][]{cha03}, SSs spinning at sub-millisecond periods should
be quite common, as accreting $\sim 0.2 M_\odot$ could spin the SS to
periods below one millisecond \citep{zhg02}.  Even evolutionary arguments
 proposed by \citet{bual01}, that can explain the non-detection of ultrafast
 radio pulsars do not apply to the detection of burst oscillations,
 which is thought to be observationally unbiased \citep[see e.g.][]{cha05}. Thus the
 observational shortage of submillisecond CSs 
 (only one out of 19 X-ray binaries that show coherent pulsations in
 their lightcurve) would be somewhat puzzling.

On the other hand if CSs are composed of nucleonic matter, the high
stiffness required to the EOS by the high measured masses can 
alone explain the observational shortage of very fast CSs:  a stiff
EOS implies that a submillisecond 
HS should both be very massive and reach the mass-shedding
regime, which seem difficult given only a few measured NS masses are
larger than 2~$M_\odot$ \citep[see e.g.][]{lapr04}. Moreover, as the
range of masses that allow a stable 
submillisecond HS is very narrow (i.e. in the case of EOS KS the
allowed mass range is $2-2.2~M_\odot$, and it is even narrower for
other EOSs, see figure \ref{fig:EOS}), this shortens further the chances we
have to observe a submillisecond HS, as a supramassive HS can either
collapse during accretion, or collapse due to braking if accretion is
stopped when the star is in the supramassive regime: in general, a
supramassive CS is expected to have a very short lifetime \citep{pa1}.
We have therefore a strong selection effect against the detection of
very fast spinning objects. Again, if there is a
transition to deconfined quark matter at very high densities in NSs, as
required by the results of HIC experiments
\citep{klaal06,grial06}, this will not change significantly this scenario.

In conclusion, we showed that when the moment of inertia of
PSR~J0737-3039A would be measured with an accuracy of $\sim 10\%$,
 we will likely be
able, by combining this measurement with the constraints presented
here, to assess the nature of CSs using only 
model-independent observational constraints. In the meanwhile, if no
other object spinning below one millisecond is found, this is probably
an argument favoring the hypothesis that CSs are composed of nucleonic matter.
Anyhow, this possibility challenges our present understanding of 
particle physics at supernuclear densities.  In fact, one should envisage a 
plausible physical mechanism to avoid the appearance of  non-nucleonic degrees of 
freedom (hyperons, quark matter, etc.) in the dense interiors of neutron stars.    


\label{lastpage}

\begin{thebibliography}{99}
\bibitem[\protect\citeauthoryear{Alford et al.}{2005}]{alfal05} Alford M. et al., 2005, \apj, 629, 969
\bibitem[\protect\citeauthoryear{Andersson et al.}{2005}]{andal05} 	
	Andersson, N., Glampedakis, K., Haskell, B., \& Watts, A. L., 2005, \mnras, 361, 1153
\bibitem[\protect\citeauthoryear{Andersson, Kokkotas \&
    Stergioulas}{Andersson et al.}{1998}]{aks98} Andersson, N.,
  Kokkotas, K., \& Stergioulas, N., 1998, 516, 307
\bibitem[\protect\citeauthoryear{Arnett \&
     Bowers}{1977}]{pasm75} Arnett, W.D., \& Bowers, R.L., 1977, \apjs,33, 415
\bibitem[\protect\citeauthoryear{Bildsten}{1998}]{bild98} Bildsten, L., 1998, \apj, 501, L89
\bibitem[\protect\citeauthoryear{Baldo, Bombaci \& Burgio}{Baldo et al.}{1997}]{bbb97} Baldo, M., Bombaci, I.,\& Burgio, G.F., 1997, \aap, 328, 274
\bibitem[\protect\citeauthoryear{Bejger \& Haensel}{2002}]{bejhe02}
  Bejger, M., \& Haensel, P., 2002,\aap, 396, 917
\bibitem[\protect\citeauthoryear{Bodmer}{1971}]{bodm} Bodmer, A. R., 1971, \prd, 4, 1601
\bibitem[\protect\citeauthoryear{Bombaci, Thampan \& Datta}{Bombaci et al.}{2000}]{bomal00} Bombaci, I, Thampan, A. V., \& Datta, V.,2000, \apj, 541, L71
\bibitem[\protect\citeauthoryear{Burderi et al.}{2001}]{bual01} Burderi L., et al., 2001, \apj, 560, L71
\bibitem[\protect\citeauthoryear{Chakrabarty et al.}{2003}]{cha03} Chakrabarty, D., Morgan,
  E. H., Muno, M. P., Galloway, D. K., Wijnands, R., van der Klis, M., \&
  Markwardt, C. B. 2003, \nat, 424, 42
\bibitem[\protect\citeauthoryear{Chakrabarty}{2005}]{cha05} Chakrabarty, D., in Rasio F.A., Stairs I.H., eds., ASP Conf. Ser., Vol. 318, Binary Radio Pulsars. Astron. Soc. Pac., San Francisco, p. 279
\bibitem[\protect\citeauthoryear{Cook, Shapiro \& Teukolsky}{Cook et al.}{1994}]{cst94} Cook G.B., Shapiro S.L and Teukolsky S.A., 1994, \apj 424, 823
\bibitem[\protect\citeauthoryear{Cottam, Paerels \& Mendez}{Cottam et
    al.}{2002}]{coal02} Cottam, J., Paerels, F., \& Mendez, M., 2002, \nat,
  420, 51
\bibitem[\protect\citeauthoryear{Datta, Thampan \&
    Bombaci}{1998}]{datal98} Datta, B., Thampan, A. V., Bombaci, I.,
  1998, \aap, 334, 943
\bibitem[\protect\citeauthoryear{Drake et al.}{2002}]{drake} Drake, J.~J., et al., 2002, \apj, 572, 996 
\bibitem[\protect\citeauthoryear{Farhi \& Jaffe}{1984}]{farhi} Farhi, E., and Jaffe, R.L., 1984, \prd 30, 2379
\bibitem[\protect\citeauthoryear{Galloway et al.}{2006}]{gal06} Galloway, D. K., et al., 2006, \apjs\, submitted (astro-ph/0608259)
\bibitem[\protect\citeauthoryear{Glendenning}{1996}]{gle96} Glendenning, N.K., 1996, ``Compact Stars'', Springer-Verlag, New York
\bibitem[\protect\citeauthoryear{Glendenning \& Moszkowki}{1991}]{glemo91} Glendenning, N.K. and Moszkowki, S.S., 1991, \prl, 67, 2414
\bibitem[\protect\citeauthoryear{Grigorian, Blaschke \&
    Kl\"ahn}{Grigorian et al.}{2006}]{grial06} Grigorian, H., Blaschke, D., \&
  Kl\"ahn, T., 2006, preprint (astro-ph/0611595)
\bibitem[\protect\citeauthoryear{Hewish et al.}{1968}]{heal68}
Hewish, A., Bell, S. J., Pilkington, J. D., Scott, P. F., Collins, R. A.,
1968, \nat, 217, 709 
\bibitem[\protect\citeauthoryear{Kaaret et al.}{2006}]{karal06} Kaaret,
  P., et al., 2006, \apjl\, submitted  (astro-ph/0611716)
\bibitem[\protect\citeauthoryear{Kl\"ahn et al.}{2006}]{klaal06}
  Kl\"ahn, T., et al., 2006, \prc, 74, 035802
\bibitem[\protect\citeauthoryear{Kl\"ahn et al.}{2006b}]{klaal06b}
  Klahn, T., et al., 2006, preprint (nucl-th/0609067)
\bibitem[\protect\citeauthoryear{Krastev \& Sammaruca}{2006}]{kraal06} Krastev, P.G., and Sammaruca, F.,  2006, \prc 74, 025808

\bibitem[\protect\citeauthoryear{Lattimer \& Prakash}{2004}]{lapr04}
  Lattimer, J.M., \& Prakash, M. 2004, Science, 204, 536

\bibitem[\protect\citeauthoryear{Lattimer \& Schutz}{2005}]{laschu05}
  Lattimer, J.M., \& Schutz, B.F., 2005, \apj, 629, 979

\bibitem[\protect\citeauthoryear{Lavagetto et al.}{2004}]{pa1} Lavagetto, G., Burderi, L.,
   D'Antona, F., Di Salvo, T., Iaria, R.,\& Robba, N.~R., 2004, \mnras, 348, 73

\bibitem[\protect\citeauthoryear{Li et al.}{1999}]{li99a} Li, X.-D., Bombaci I., Dey M., Dey J, \& van den Heuvel E.P.J., 1999, \prl, 83, 3776


\bibitem[\protect\citeauthoryear{Lugones \& Horvath}{2003}]{lug03}
  Lugones, G., \& Horvath, J.E., 2003,\aap, 403, 173

\bibitem[\protect\citeauthoryear{Nice et al.}{2005}]{nial05} Nice,
  D.,Splaver, E. M., Stairs, I. H., L\"ohmer, O.; Jessner, A.; Kramer,
  M.; Cordes, J. M., 2005, \apj, 634, 1242	

\bibitem[\protect\citeauthoryear{Prakash et al.}{1997}]{prak97}  
Prakash, M., Bombaci, I., Prakash, M., Ellis, P.J.,  Lattimer, J.M., \&   Knorren, R. 1997,  Phys. Rep. 280, 1  

\bibitem[\protect\citeauthoryear{Ransom et al.}{2005}]{raal05} Ransom,
  S., Hessels, J.W.T., Stairs, I. H., Freire, P.C.C, Camilo, F.,
  Kaspi, V. M., \& Kaplan, D. L. 2005, Science,307, 892
\bibitem[\protect\citeauthoryear{Salgado et al.}{1994}]{saal94}
  Salgado, M., Bonazzola, S., Gourgoulhon, E., Haensel, P., 1994,
  \aap, 291, 155
\bibitem[\protect\citeauthoryear{Thorsett \& Chakrabarty}{1999}]{tocha99}   Thorsett, S.E., \&
  Chakrabarty, D., 1999, \apj,  512, 288
\bibitem[\protect\citeauthoryear{Tr\"umper et al.}{2004}]{trumper04} Tr\"umper, J.~E., 
Burwitz, V., Haberl, F., \& Zavlin, V.~E., 2004, Nucl. Phys. B 
Proc. Suppl., 132, 560
\bibitem[\protect\citeauthoryear{van der Klis}{2006}]{vdk06} van der Klis, M., 2006,
  in Lewin W.H.G., van der Klis M., eds, Compact Stellar X-ray
  sources. Cambridge University Press, Cambridge, p. 39
\bibitem[\protect\citeauthoryear{Wijnands et al.}{2003}]{wijal03} Wijnands, R.,
  van der Klis, M., Homan, J., Chakrabarty, D., Markwardt, C. B., \&
  Morgan, E. H., 2003, \nat, 424, 44
\bibitem[\protect\citeauthoryear{Witten}{1984}]{witt} Witten, E. 1984, \prc, 30, 272
\bibitem[\protect\citeauthoryear{Zdunik, Haensel \&
    Gourgoulhon}{Zdunik et al.}{2002}]{zhg02} Zdunik, J. L., Haensel,
  P., \& Gourgoulhon, E., 2002, \aap, 381, 933
\bibitem[\protect\citeauthoryear{Zhang et al.}{2006}]{zaal06}
  Zhang, C. M., et al., 2006, \mnras\, in press, (astro-ph/0611659)
\end{thebibliography}
\end{document}